\documentstyle[sprocl]{article}
\input psfig %bibliographystyle{unsrt}    
\def\Journal#1#2#3#4{{#1} {\bf #2}, #3 (#4)}
\def\ApJ{\em Ap. J.}

\def\PRD{{\em Phys. Rev.} D}

\def\be{\begin{equation}}
\def\ee{\end{equation}}
\def\bea{\begin{eqnarray}}
\def\eea{\end{eqnarray}}

\begin{document}
\title{INTRODUCTION \& OVERVIEW: CMB SESSIONS}
\author{ G.F. SMOOT }
\address{Lawrence Berkeley National Laboratory,\\
University of California, Berkeley CA 94720, USA}
\maketitle\abstracts{
This is a very exciting time for the CMB field. It is widely recognized 
that precision measurements of the CMB can provide a definitive test 
of cosmological models and determine their parameters accurately. 
At present observations give us the first rough results but ongoing 
experiments promise new and improved results soon and eventually 
satellite missions (MAP and COBRAS/SAMBA now named Planck) are expected to 
provide the requisite precision measurements. Other areas such as 
observations of the spectrum and Sunyaev-Zeldovich effect are also 
making significant progress.}
\section{Introduction}
There has long been anticipation that cosmic microwave background 
(CMB) radiation would provide significant information about the 
early Universe due to its early central role and its general lack of 
interaction in the later epochs. 
\subsection{COBE}\label{subsec:cobe}
Though there have been many observations of the CMB since its 
discovery by Penzias and Wilson \cite{Penzias65} in 1964, 
the Cosmic Background Explorer satellite,  COBE, 
provided two watershed observations. 
The first key observation is that the CMB is extremely well described 
by a black-body spectrum \cite{Mather94,Fixsen96}. 
This observation of the CMB thermal origin strongly affirms the hot 
Big Bang predictions and tightly constrains possible energy releases, 
ruling out explosive and other exotic structure formation 
scenarios. 
The spectral measurement of the temperature as a blackbody and 
the dipole anisotropy as its derivative provides the basis of 
knowledge of the spectral shape of the higher order CMB anisotropies 
and thus a means to separate them from the foregrounds.

The detection of primordial CMB anisotropies \cite{Smoot92} is
the second key from COBE. 
The COBE large angular scale map of the microwave sky and detection of 
intrinsic anisotropies provides support for the gravitational 
instability picture and thus a link to large scale structure, an anchor 
point in the magnitude of fluctuations, an impetus and 
guidance to the field as a whole.

\subsection{Theory}\label{subsec:theory}
In addition to the increased attention to observations and the 
development of experimental techniques, a major thrust in the field 
has been the improvement in theory and ideas for extracting information 
from the data. 
Theoretical work has provided ideas and means to calculate the 
observable effects of various cosmological scenarios from 
standard Cold Dark Matter (sCDM) and its variants, various 
inflationary models through to nearly a good description of 
topological defects (an area still in active development). 
Other work has lead to a better physical intuition of the mechanisms 
involved \cite{Hu,Albrecht, Others}.

The inverse problem of extracting the scenario and the appropriate 
cosmological parameters including error estimates is an active area 
following the pioneering paper by Llyod Knox \cite{Knox95} and a 
seminal paper by Jungman et al.\cite{Jungman}. 
The understanding that one can extract cosmological parameters with 
accuracy is now driving the excitement in the field in equal measure 
with the knowledge that the CMB anisotropies are there and are 
observable.

\section{Current Observations}
Since the COBE DMR detection of anisotropy, 
over a dozen groups have reported anisotropy detections and 
some interesting upper limits. 
The current power spectrum observations are summarized in 
the left panel of Figure 1.
\begin{figure} %\rule{5cm}{0.2mm}\hfill\rule{5cm}{0.2mm} 
%\vskip 2.5cm \rule{5cm}{0.2mm}\hfill\rule{5cm}{0.2mm}
\psfig{figure=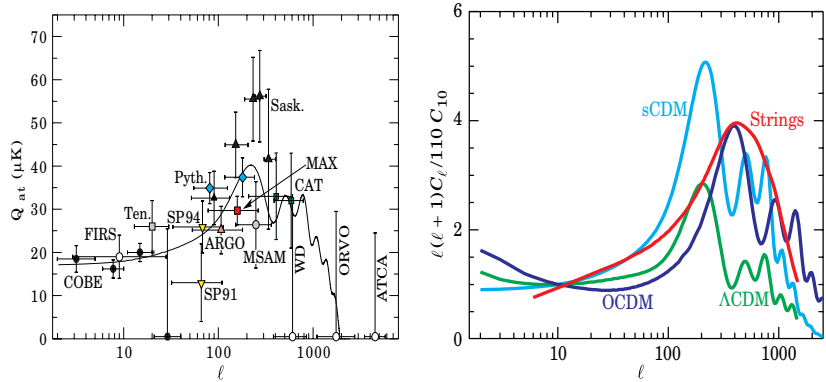,height=2in} 
\caption{{\bf Left Panel:} Current CMB anisotropy power spectrum observations. 
The horizontal axis $\ell$ is Legrendre polynomial number and 
is inversely proportional to the angular scale. 
The vertical axis is the RMS (rather than power) temperature 
variation for that angular scale.
The solid curve is the calculated anisotropy for sCDM. \hfill \hfill  
{\bf Right Panel:} Theoretical CMB anisotropy power spectrum 
for various models. 
%The horizontal axis $\ell$ is Legrendre polynomial number and 
%is inversely proportional to the angular scale. 
The vertical axis is the temperature anisotropy power for that 
angular scale.}
\end{figure}

For comparison the right panel shows the predicted anisotropy power 
spectrum for a number of different cosmological models.

\section{Future and CMB Sessions}
The CMB field is extremely active and exciting precisely because of 
the combination of rapid observational and theoretical development 
with definitive space missions in the coming decade and 
of the expectation that those observations will provide us with 
accurate determination of cosmological parameters.
The great progress and interest are shown by the coverage in the 
invited talks by Scott Dodelson \cite{Dodelson}, Lyman 
Page \cite{Page}, and Neil Turok \cite{Turok} and in the talks in these 
two CMB sessions as well as posters and discussions by the meeting 
attendees.

A large total effort is necessary 
to achieve these lofty aims. 
The many people in the field must work together in a combination of 
competition and collaboration. 
Collaboration is necessary because the tasks are large and difficult as 
well as subtle.
Competition is necessary as many people, especially the large 
number of young and very excellent scientists, must continue 
to establish their careers and accomplishments and 
because some competition is useful to keep eveyone on their toes.
However, it is important to keep in mind that our ultimate goal is 
the accurate testing and determination of cosmological models and 
parameters and that transcends any one group. 
The total resources that are and will be allocated to this effort 
are temendous - both in financial and human capital terms.
It is everone's responsibility to work towards this ultimate goal 
and especially for the senior scientists to promote this and set a tone 
of friendly and helpful competitive cooperation.
The rapid open sharing of data is a good standard but more is needed.

In addition to this cooperation a number of things must fall into place.
The first is outstanding instrumentation and techniques for making 
the observations. 
Quality instrumentation is particularly important for the forthcoming 
satellite missions as they are costly in terms of money and mission opportunity.

The next is an understanding of how to process the data and 
then to turn the calibrated data into maps, power 
spectra, and other useful forms. 
This is intimately linked to understanding both the instruments and the 
foregrounds: galactic and extragalactic including the SZ effect.
Work is needed in this area and has begun.

Finally the calibrated separated data must be used in the inverse 
problem to test cosmological models and recover their best-fitted 
parameters. 

All of these efforts require a significant advancement in their technology.  
The talk by John Carlstrom \cite{Carlstrom} on making Sunyaev-Zeldovich maps 
using new technology HEMT (high electron mobility transistor) 
amplifiers is an example of the advances that can be made in a field 
with new techniques/technologies.
We have only recently come to appreciate the computorial 
complexity of utilizing a million-plus pixel map.  
Already a number of approaches are being tried and programs being 
developed to address these issues.

The existence of already ongoing programs, e.g. the balloon-borne 
instruments: BOOMERANG/MAXIMA, MSAM/TOPHAT, 
HACME/BEAST, 
and the interferometers: VCA, CBI, VSA,
will provide additional and appropriate data to test these techniques 
in a very short time horizon. 
These plus the pressure from the future missions and observations
provide us with both an exciting and challenging field.

{\it Acknowledgments}
This work is supported in part by the Director, Office of
Energy Research, Office of High Energy and Nuclear Physics, Division of
High Energy Physics of the U.S. Department of Energy under Contract No.
DE-AC03-76SF00098.

\end{document}